\documentclass[conference]{IEEEtran}
\usepackage[T1]{fontenc}% optional T1 font encoding
\usepackage{amsmath}
\usepackage[cmintegrals]{newtxmath}
\usepackage{url}
\usepackage{soul,color,graphicx,float,epstopdf}
\usepackage{tablefootnote,textcomp,gensymb}
\usepackage{eurosym,booktabs,multirow}
\usepackage[caption=false]{subfig}
\usepackage{amsfonts} 
\usepackage{algorithmic}
\usepackage{array}
\usepackage{stfloats}
\usepackage{float}
\usepackage{mathtools}
\usepackage{graphicx}
\usepackage{pdfpages}
\usepackage{subfig} 
\usepackage{balance}
\usepackage{xcolor}
\usepackage{comment}
\usepackage[normalem]{ulem}
\usepackage{dsfont}
\usepackage[linesnumbered, ruled]{algorithm2e}
\usepackage{enumitem}
\usepackage{tikz}
\usetikzlibrary{arrows,shapes,positioning,shadows,trees}

\usepackage{verbatim}
\usepackage{cite}

\definecolor{Orange}{rgb}{1,0.5,0}
\newcommand{\eqr}[1]{$#1$}

\newcommand*{\rom}[1]{\expandafter\@slowromancap\romannumeral #1@}

\usepackage{xcolor}
\usepackage{balance}

\begin{document}
\title{DRL-Based Sidelobe Suppression for Multi-focus Reconfigurable Intelligent Surface}

\author{\IEEEauthorblockN{
Wei Wang\IEEEauthorrefmark{1},
Peizheng Li\IEEEauthorrefmark{2},
Angela Doufexi\IEEEauthorrefmark{1},
Mark A Beach\IEEEauthorrefmark{1}
}                                     % ...

\IEEEauthorblockA{
\IEEEauthorrefmark{1}% 1st affiliations
Department of Electrical and Electronic Engineering, University of Bristol, U.K.\\
\IEEEauthorrefmark{2}% 2nd affiliations
Bristol Research \& Innovation Laboratory, Toshiba Europe Ltd., U.K.\\
\{wei.wang, A.Doufexi, M.A.Beach\}@bristol.ac.uk, peizheng.li@toshiba-bril.com}}
% \IEEEauthorblockA{\IEEEauthorrefmark{3}% 3rd affiliations
% (Affiliation): dept. name of organization, name/acronyms of organization, City, Country, e-mail address*}
% \IEEEauthorblockA{\IEEEauthorrefmark{4}% 4th affiliations
% (Affiliation): dept. name of organization, name/acronyms of organization, City, Country,
%  e-mail address*}  
 % \IEEEauthorblockA{ \emph{*at least one e-mail address should be indicated above} }



\maketitle

\begin{abstract}
Reconfigurable intelligent surface (RIS) technology is receiving significant attention as a key enabling technology for 6G communications, with much attention given to coverage infill and wireless power transfer. However, relatively little attention has been paid to the radiation pattern fidelity, for example, sidelobe suppression. When considering multi-user coverage infill, direct beam pattern synthesis using superposition can result in undesirable sidelobe levels. To address this issue, this paper introduces and applies deep reinforcement learning (DRL) as a means to optimize the far-field pattern, offering a 4dB reduction in the unwanted sidelobe levels, thereby improving energy efficiency and decreasing the co-channel interference levels.
\end{abstract}

\vskip0.5\baselineskip

\vspace{-1.00mm}
\section{Introduction}
\label{sec:intro}
% \vspace{-1.00mm}
Reconfigurable Intelligent Surfaces (RIS) are expected to play a crucial role in the 6G network~\cite{di2020smart} for its ability to adjust the physical wireless propagation environment~\cite{basar2019wireless}.
As the fundamental component of RIS, the reconfigurable reflectarray (RRA)~\cite{liu2017concepts} consists of a number of passive elements eliminating the need for a radio frequency (RF) chain to modem and re-radiate the waveform. The electronically tunable RRA elements offer passive beamforming, bringing the benefit of low fabrication cost and high energy efficiency~\cite{huang2019reconfigurable}. Meanwhile, the large array compensates for its low reflection efficiency from the passivity, outperforming the active relays~\cite{bjornson2019intelligent}. Therefore, RIS is treated as an ideal solution to decrease the density of 6G base stations (BS) deployment.

Recently, several works~\cite{yang2016programmable,taghvaee2022enabling, tran2022multifocus} have investigated the multi-reflection ability of RIS regarding its potential applications in broadcasting~\cite{taghvaee2022enabling} and wireless power transformation~\cite{tran2022multifocus}. With the demand for RIS to support scenarios with geographically distributed users, the creation of highly directional multi-focus beams emerges as a main challenge in the process of RRA reconfiguration. The passive RRA generates reconfigurable reflections through electronically controlled phase and amplitude of each element~\cite{wu2019towards}. As the main focus of this paper, the combined phase and amplitude configuration set within the RRA is termed the \textbf{\textit{reflection profile}}.

Typically, a multi-focus reflection profile forms by overlaying single-focus profiles, as each single-focus sub-profile reflects in a specific direction, enabling the RIS for a synchronous multi-focus reflection. However, superimposing these sub-profiles introduces interference, resulting in the pollution of each sub-profile and consequently disrupting passive sub-beams based on the profiles.
As a result, non-negligible sidelobes emerge alongside intended beams due to accumulative errors in the periodical arrays. That leads to an increase in co-channel interference and the energy efficiency decay of the entire network.
This issue has been discussed in a few works~\cite{tran2022multifocus, nayeri2013design}, but the involved reflectarray works merely in the near-field of the feed as a part of the transmitter, and their proposed global search optimizers are also not feasible enough due to the high time consumption. Therefore, in RIS applications where the RRA intercepts a far-field parallel incidence from the BS and disperses it across multiple directions, a more efficient approach to suppress undesired sidelobes becomes imperative for the optimal reflection profile of the multi-focus RRA.

As an emerging technique, deep reinforcement learning (DRL) has demonstrated its effectiveness in various real-world challenges, including RIS~\cite{alexandropoulos2022pervasive, huang2020reconfigurable, huang2021multi, wang2023drl}. DRL excels in sequential decision-making, enabling agents to learn from experiences and improve their performance progressively. In addressing the sidelobe suppression issue of the multi-focus RIS reflection, if the sidelobe generation mechanism can be structured as a sequential decision-making problem and the relevant influence parameters can be explored by the DRL agent, DRL would be a more potent alternative compared to other optimization approaches~\cite{tran2022multifocus},~\cite{nayeri2013design}. This work centers on optimizing the RIS reflection profile through DRL, aiming to bolster the main lobe reflections while attenuating sidelobe gains.

% \noindent
\textit{Contributions:} 
This paper analyzes the generation mechanism of intense sidelobes in RIS multi-reflection, and proposes the randomizing parameter $\delta$ and suppression parameter $\xi$ which are crucial for sidelobe suppression.
% This paper leverages antenna array theory to approximate RIS reflection, focusing on far-field incident waves. By delving into the generation mechanism of intense sidelobes in RIS multi-reflection, we propose the randomizing parameter $\delta$ and suppression parameter $\xi$ which are crucial for sidelobe suppression. 
Meanwhile, a novel DRL algorithm is designed to find the optimal combination of these parameters. To the best of our knowledge, this is the first work to address the high-intensity sidelobes in RIS multi-reflection with a far-field parallel incidence. 
% Notably, we pioneer the use of DRL in improving reflections.
The entire analysis is conducted on an RRA simulation producing reflect beams aiming at two directions. However, we envision our approach's adaptability to more complicated scenarios to broaden its potency.

\textit{Paper outline:} This paper is constructed as follows. Sec.~\ref{subsec:Fundamental Theory} introduces the antenna array theory and the RIS-reflection approximation. Then, Sec.~\ref{subsec:Problem Formulation} details the problem formulation of sidelobe suppression to an extreme value problem. Sec.~\ref{sec:Sidelobe Suppression Method} and \ref{sec:DRL-Based Algorithm} elaborate on the proposed sidelobe suppression method and the DRL algorithm, respectively. Finally, Sec.~\ref{sec:Result and Discussion} displays and discusses the simulation results, and Sec.~\ref{sec:Conclusion} concludes this paper.

%%%%%%%%%%%%%%%%%%%%%%%%%%%%%%%%%%%%%%%%%%%%%%%%%%%%%%%%%%%%%%%%%%%%%%%%%%%%%
\vspace{-1.00mm}
\section{Preliminary and Problem Formulation}
\label{sec:Fundamental Theory and Problem Formulation}
% \vspace{-1.00mm}
\subsection{Fundamental Theory}
\label{subsec:Fundamental Theory}
\begin{figure}
\vspace{-1.25mm}
\centering
\includegraphics[width=65mm]{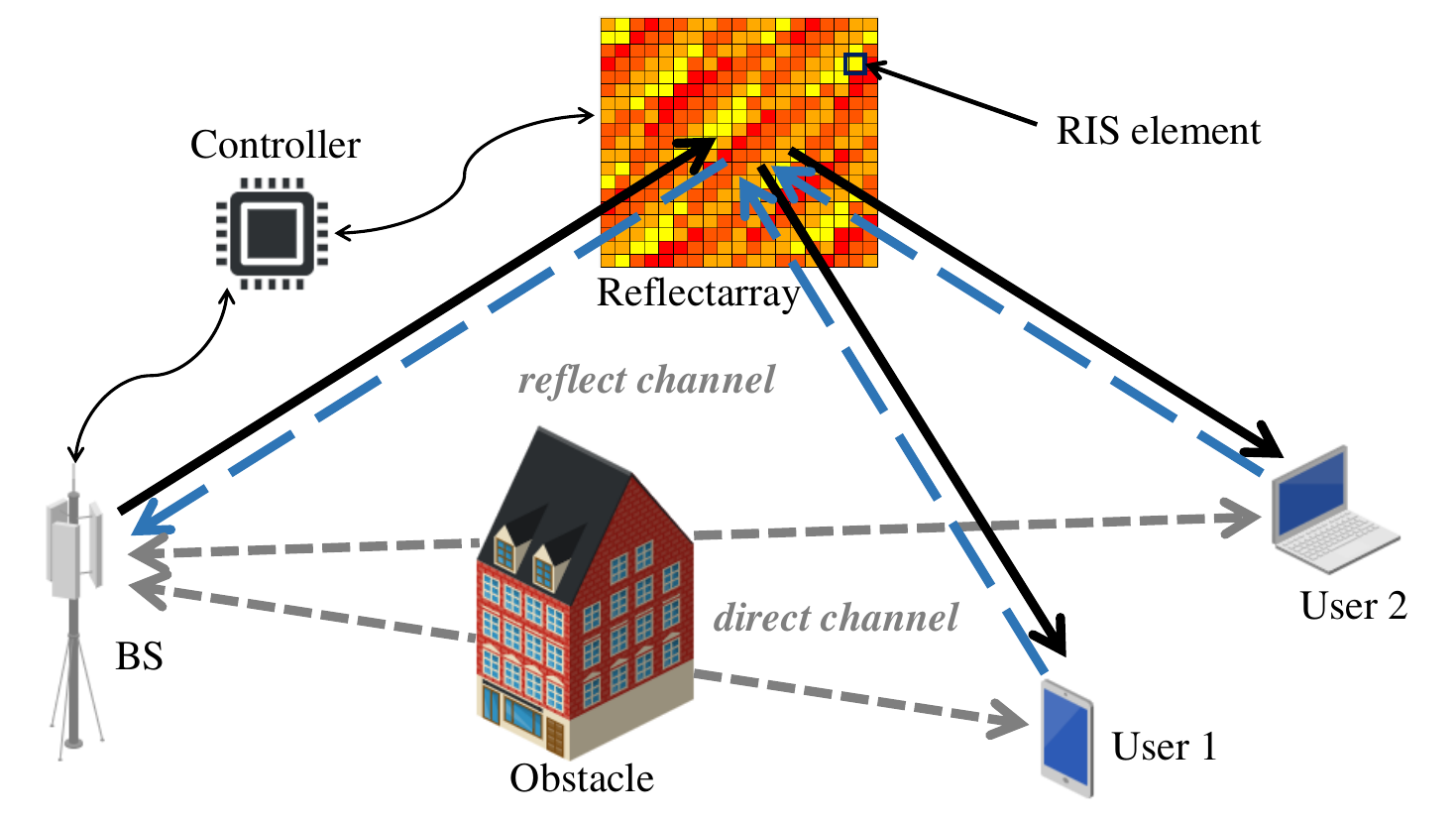}
\vspace{-3.25mm}
\caption{The components and the application scenario of a RIS system. Its direct channels are blocked by obstacles.}
\label{fig:RIS}
\vspace{-4.50mm}
\end{figure}

Fig.~\ref{fig:RIS} illustrates RIS operational principles within a deployment scenario. A controller connects to the BS, determining the Angle of Arrival (AoA) and Angle of Departure (AoD) of the wave, then computing the necessary reconfigured profile for the RRA to realize the desired reflections.
Considering a typical RIS-assisted cascaded channel with \textit{N}-element BS, \textit{M}-element RRA, and \textit{K} single-antenna user equipment (UE), the received signal at \textit{k}-th UE $y_{k}$ can be expressed as:
\begin{equation}
    y_{k} = (\boldsymbol{\rm{h}}_{r,k} \rm{diag}(\boldsymbol{\Theta})\it{\boldsymbol{\rm{G}}}) \boldsymbol{\rm{s}}^{\rm{T}} + \it{w}_{\emph{k}}\,,
\label{equi:1}
\end{equation}
where $\boldsymbol{\rm{s}} \in \mathbb{C}^{N}$ is the signal transmitted from BS, $\it{w}_{\emph{k}} \in \mathbb{C}$ is the received noise with zero mean and variance $\sigma^{2}$ at \textit{k}-th UE. $\boldsymbol{\rm{G}} \in \mathbb{C}^{M\times N}$ and $\boldsymbol{\rm{h}}_{r,k} \in \mathbb{C}^{M}$ denote the BS-RIS and RIS-UEs channel respectively. In (\ref{equi:1}), $\rm{diag}(\boldsymbol{\Theta})$ indicates a diagonal matrix with each element $\beta_{m}e^{j\theta_{m}}$ denoting the amplitude response~$\beta_{m}$ and phase shift~$\theta_{m}$ on \textit{m}-th reflectarray element. Under different controlling voltages, the reflection amplitude~$\beta$ and phase~$\theta$ of each reflectarray element can be adjusted accordingly. To simplify the design, $\beta$ and $\theta$ are normally quantified to several discrete levels. 
RIS employs controllable $\beta$ and $\theta$ parameters to generate adaptable RRA reflect profiles, enabling the passive beamforming implementation.

RRAs come in three main types: (1) phase-only array~\cite{pei2021ris, araghi2022reconfigurable}, (2) amplitude-only array~\cite{cao2021design}, and (3) amplitude-phase-joint array~\cite{yang2016design, wang2022broadband}, wherein the phase-only array is often the most achievable~\cite{wang2022broadband}, and the amplitude response ($\beta$) remains consistent and constant across all elements. Consequently, this type relies solely on variable phases for reconfigurable reflection. By integrating a suitable varactor diode onto the RRA element and adjusting its voltage, multi-bit dynamic reflection phases can be realized. In this paper, we focus on the phase-only RRA for reflection optimization as well.
\begin{figure}
\vspace{-1.25mm}
    \subfloat[\label{fig:phase}]{
      \begin{minipage}[t]{0.4\linewidth}
        \centering 
        \includegraphics[width=1.5in]{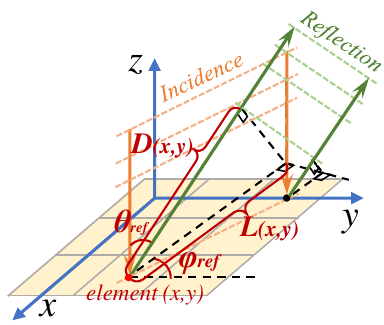}   
      \end{minipage}%
      }
      \hfill
        \subfloat[\label{fig:OPD}]{
      \begin{minipage}[t]{0.5\linewidth}   
        \centering   
        \includegraphics[width=1.65in]{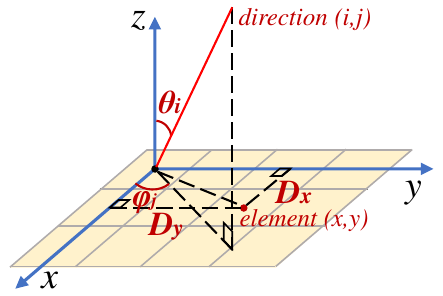}   
      \end{minipage} 
      }
      \vspace{-1.25mm}
      \caption{The schematic diagrams of (a) phase difference calculation and (b) reflection pattern calculation.
      } \label{fig:diagram}
      \vspace{-4.50mm}
\end{figure}

In the scenario shown in Fig.~\ref{fig:phase} that a perpendicular incident plane wave being reflected in the single direction [$\theta_{ref}, \varphi_{ref}$] by the phase-only RRA, the phase compensation $\it{\Delta}_{(x,y)}$ that the element at location~[\textit{x},\textit{y}] should achieve can be denoted as:
\begin{equation}
    D_{(x,y)}=L_{(x,y)}\cdot \rm{sin}\theta_{ref}\,,
\end{equation}
\begin{equation}
    \it{\Delta}_{(x,y)}=\it{k}\cdot D_{(x,y)}\,,
\label{equi:2}
\end{equation}
where $L_{(x,y)}$ is the distance from the RRA's edge to the \textit{m}-th RRA element along the reflected direction, while $D_{(x,y)}$ represents the reflection optical path difference calculated from $L_{(x,y)}$. $\textit{k}$ indicates the wavenumber, defined as $2\pi/\lambda$, with $\lambda$ being the wavelength of the incident signals in vacuum. Upon achieving the single-focus reflection profile \eqr{P_{sing}}, the multi-focus reflection profile \eqr{P_{mul}} can be composed by overlaying several single-focus profiles. This profile calculation method can also be extended to non-perpendicular incident plane waves by incorporating the incident phase difference, calculable in a similar manner.

Fig.~\ref{fig:OPD} shows the schematic diagram of the far-field reflection pattern calculation, in which the optical path difference (OPD) in the direction~[\textit{i},\textit{j}] of the RRA element at location [\textit{x},\textit{y}] can be expressed as:
\begin{equation}
    \it{OPD}_{(x,y),(i,j)}=D_{x}\cdot \cos(\varphi_{j})\sin(\theta_{i})+D_{y}\cdot \sin(\varphi_{j})\sin(\theta_{i})\,,
\end{equation}
where $D_x$ and $D_y$ are the distances between the element~[\textit{x},\textit{y}] and the array edge element in the x- and y-directions respectively. Then, the far-field~\textbf{E} of the RRA is easy to be achieved by:
\begin{equation}
    % \it{E}=\it{E}_{single}\cdot\sum_{x=0}^{X-1} \sum_{y=0}^{Y-1} \beta_{(x,y)}\cdot e^{j[k\cdot OPD_{(x,y),(i,j)}+\theta_{(x,y)}]}
    \it{\boldsymbol{\rm{E}}}=\it{\boldsymbol{\rm{E}}}_{single}\cdot\sum_{x={\rm 0}}^{X-{\rm 1}} \sum_{y={\rm 0}}^{Y-{\rm 1}} \beta_{(x,y)}\cdot e^{j[k\cdot OPD_{(x,y),(i,j)}+\Delta^{*}_{(x,y)}]}\,,
\end{equation}
where $\boldsymbol{\rm{E}}_{single}$ represents the far-field of a single element and here it is assumed to be isotropic for simplified analysis. $\it{\Delta}^{*}_{(x,y)}$~is the actual phase compensation of the element at [\textit{x},\textit{y}] quantized from $\it{\Delta}_{(x,y)}$ in (\ref{equi:2}), and is depending on the phase resolution of the RRA.

\subsection{Problem Formulation}
\label{subsec:Problem Formulation}
Based on the aforementioned theory, the far-field pattern of the reflectarray can be computed for diverse reconfigurable profiles, encompassing both single-focus and multi-focus reflections. Given the substantial RRA array size, some unwanted sidelobes will be overlapped to a high intensity in the far-field beside the desired beams. Here, we define the far-field intensity of the intended beams as:
\begin{subequations}
\begin{equation}
    \it{E_{intended}}=\min\{\mathop{\max\{\it{\boldsymbol{\rm{E}}_{\rm 1}}\}}\limits_{(\theta_{ref},\varphi_{ref})\in{S_{\rm 1}}},\cdots,\mathop{\max\{\it{\boldsymbol{\rm{E}}_{T}}\}}\limits_{(\theta_{ref},\varphi_{ref})\in{S_T}}\}
\end{equation}
\begin{equation}
    \it{s.t.}\quad \mathop{\it{S_t}\colon (\theta-\theta_{ref}^t)^{\rm 2}+(\varphi-\varphi_{ref}^t)^{\rm 2}\leq R^{\rm 2}}\limits_{\it{t={\rm 1},{\rm 2},\cdots,T}}\,.
\end{equation}
\end{subequations}

Similarly, the far-field intensity of all unwanted sidelobes is defined as:
\begin{subequations}
\begin{equation}
    \it{E_{unwanted}}=\mathop{\max\{\it{\boldsymbol{\rm{E}}^{*}}\}}\limits_{(\theta_{ref},\varphi_{ref})\in{S^*}}
\end{equation}
\begin{equation}
    \it{s.t.}\quad \mathop{\it{S^*}\colon (\theta-\theta_{ref}^t)^{\rm 2}+(\varphi-\varphi_{ref}^t)^{\rm 2}> R^{\rm 2}}\limits_{\it{t={\rm 1},{\rm 2},\cdots,T}}\,,
\end{equation}
\end{subequations}
where $\it{T}$ is the number of desired reflection beams, $\it{S_t(t={\rm 1},{\rm 2},\cdots,T)}$ indicates the circle with the center point at $(\theta_{ref}^t,\varphi_{ref}^t)$ and a radius of $\it{R}$ (10 degrees as experience value). It is employed to pinpoint beams, considering potential imprecision in real far-field reflections.
In Cartesian coordinates, $(\theta_{ref}^t,\varphi_{ref}^t)$ indicates the direction of each expected reflection beam. Hence,  the $E$-field's maximum value within each circle corresponds to the far-field intensity reflected by the RRA, and the minimum of them represents the lowest intensity across all the intended beams. Simultaneously, the maximized $E$ of the space outside of $\it{S_t}$ is the highest intensity of all undesired sidelobes.

\begin{figure}
\vspace{-1.25mm}
\centering
\includegraphics[width=90mm]{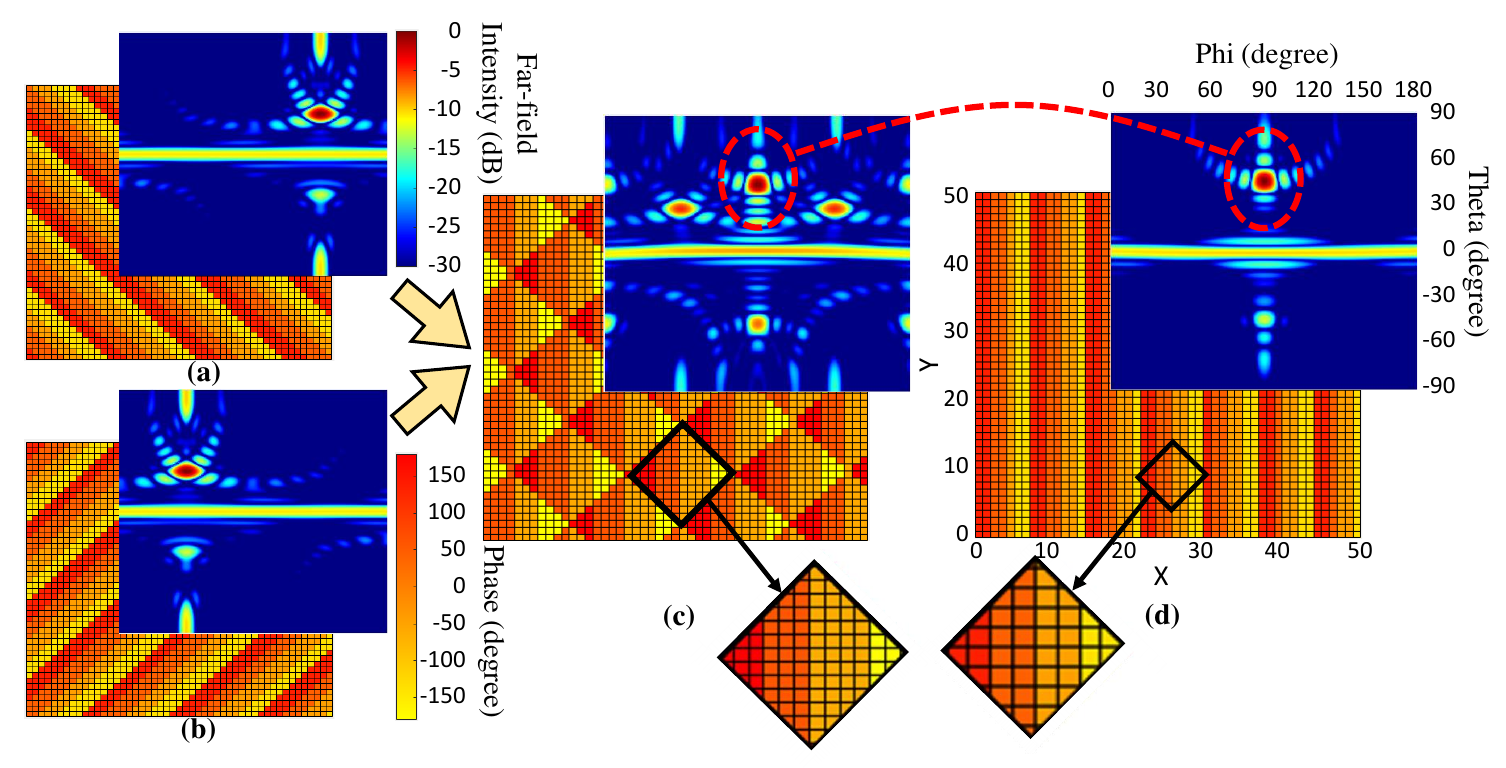}
\vspace{-6.25mm}
\caption{The generation mechanism of the unwanted sidelobes with high intensity. (a) Single-focus profile and its far-field pattern of ($\theta = 30\degree, \varphi = 135\degree$), (b) single-focus profile and its far-field pattern of ($\theta = 30\degree, \varphi = 45\degree$), (c) the overlapped multi-focus profile and its far-field pattern, (d) single-focus profile and its far-field pattern of the unwanted sidelobe in (c).}
\label{fig:mechanism}
\vspace{-4.50mm}
\end{figure}

In this paper, we aim to maximize the difference between $\it{E_{intended}}$ and $\it{E_{unwanted}}$ by optimizing the multi-focus reflection profiles. Accordingly, the optimization problem is given as:
\begin{equation}
    P:\quad \max\{\it{E_{intended}}-\it{E_{unwanted}}\}\,.
\end{equation}

% It is known that the quantized phase shifts also have a significant effect on the sidelobes’ generation~\cite{yang2016study}, and a higher resolution reflectarray hardware will improve them a lot. However, the high resolution asks for more accurate phase shift quantizations for the RRA element, which will increase the complexity and the cost of the entire RIS system.

%%%%%%%%%%%%%%%%%%%%%%%%%%%%%%%%%%%%%%%%%%%%%%%%%%%%%%%%%%%%%%%%%%%%%%%%%%%%%
\vspace{-1.50mm}
\section{Sidelobe Suppression Method}
\label{sec:Sidelobe Suppression Method}
% \vspace{-1.50mm}
When attempting to directly overlap two single-focus profiles into a multi-focus profile, undesired high-intensity sidelobes emerge alongside the intended beams, as seen in Fig.~\ref{fig:mechanism}~(a)-(c). This occurrence can be attributed to the inherent tendency of the overlapped multi-focus profile to reflect in additional directions. This is evident in Fig.~\ref{fig:mechanism}(c), where the rhombus section in the overlapped profile strongly inclines to reflect in the $\varphi = 90\degree$ direction.

To verify the aforementioned conjecture, as illustrated in Fig.~\ref{fig:mechanism}(d), a single-focus profile is created to reflect in $(\theta = 45\degree, \varphi = 90\degree)$, as the most prominent undesired sidelobe observed in Fig.~\ref{fig:mechanism}(c). The calculated far-field pattern reveals the high-intensity lobes at ($45\degree, 90\degree$), ($-45\degree, 90\degree$), and the incident direction $\theta = 0\degree$. These high-intensity lobes also manifest in Fig.~\ref{fig:mechanism}(c) alongside the intended beams. Both Fig.~\ref{fig:mechanism}(c) and Fig.~\ref{fig:mechanism}(d) exhibit a rhombus part with the same reflection tendency, which intensifies an undesired beam due to the big array size. To tackle these elevated sidelobes in multi-reflection RIS, the undesirable sidelobe tendencies from overlapped multi-profile need elimination, along with the disruption of periodicity. Therefore, in this paper, we propose a DRL-based method as well as two particular parameters, $\delta$ and $\xi$, are summarised for the sidelobe suppression.

\begin{figure}
\vspace{-1.25mm}
\centering
\includegraphics[width=90mm]{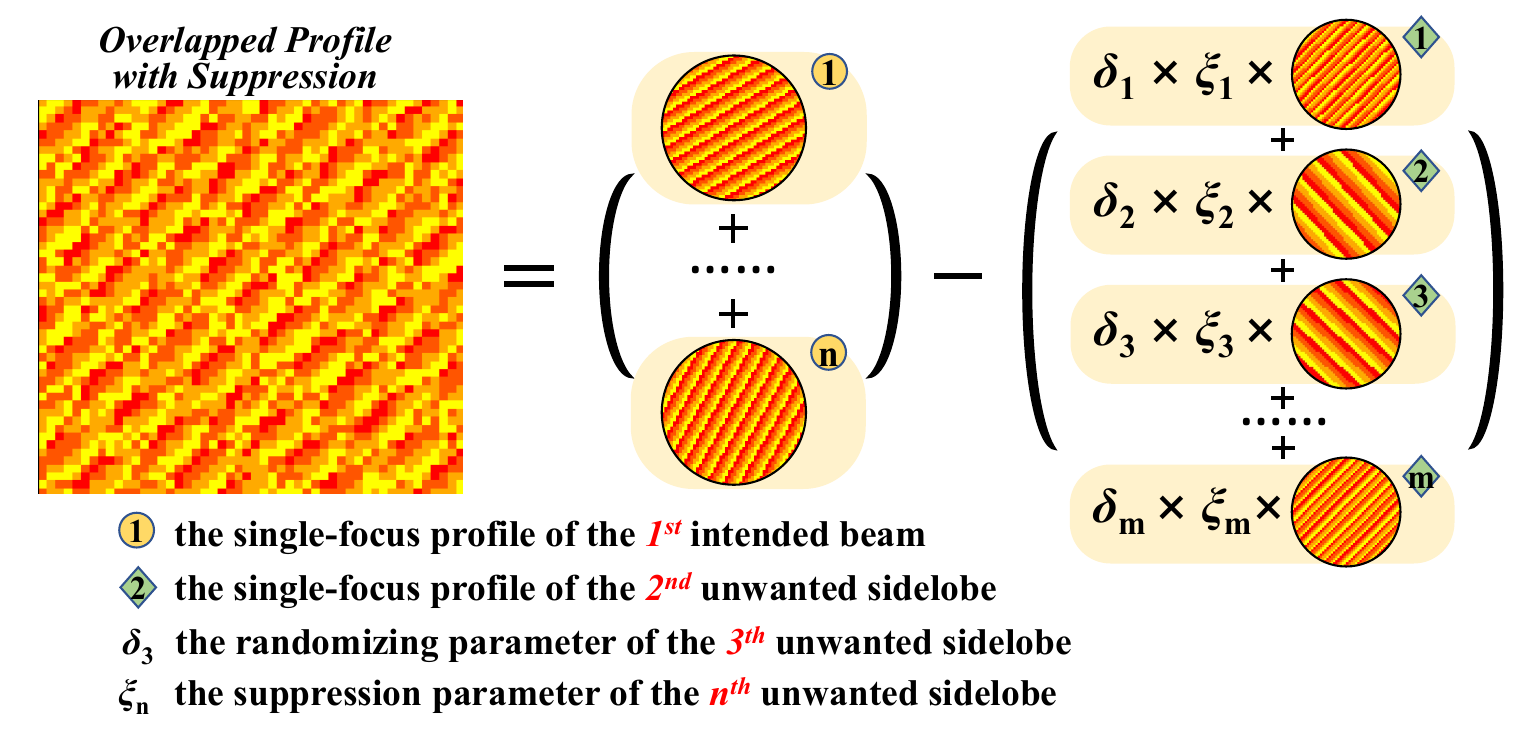}
\vspace{-6.25mm}
\caption{The sidelobe suppression method, in which the single-focus profile of each intended beam should be added up and then the component of each sidelobe should be removed for suppression.}
\label{fig:method}
\vspace{-4.50mm}
\end{figure}

Algorithm~\ref{al:DRL} outlines the proposed method. Initially, we compute the far-field pattern of the unsuppressed superposition to identify high-intensity sidelobes. A global search algorithm determines their directions, and their single-focus profiles are derived based on incidence and sidelobe directions. After that, in Fig.~\ref{fig:method}, we introduce two critical parameters for sidelobe suppression: \textit{the random parameter} $\boldsymbol{\delta}$ generates a random sequence to eliminate the values of some elements in each single-focus profile canceling the periodicity of the overlapped profile; \textit{the suppression parameter} $\boldsymbol{\xi}$ controls the intensity of each suppressed beam. Adjusting these parameters removes undesired reflection components from high-intensity sidelobes, enhancing the disparity between desired and undesired far-field reflections. Here, we search for the optimal parameter pair ($\delta$ and $\xi$) using DRL\footnote{The DRL training principle has been well-explained in a plethora of literature. Readers could refer to~\cite{sutton2018reinforcement} for a better understanding.}.
%%%%%%%%%%%%%%%%%%%%%%%%%%%%%%%%%%%%%%%%%%%%%%%%%%%%%%%%%%%%%%%%%%%%%%%%%%%%%
\vspace{-1.50mm}
\section{DRL-Based Algorithm}
\label{sec:DRL-Based Algorithm}
% \vspace{-1.50mm}

Due to the complexity of theoretically calculating optimal suppression parameters, we employ a deep Q-network (DQN) to establish the mapping between the RIS state and parameters guided by performance indicators. In the DRL environment, we construct far-field radiation based on the quantized phase shift levels of each RRA element. Actions are generated to adjust the overlapped RRA reflection profile, enabling the calculation of the radiation beam pattern. Subsequently, the performance of sidelobe suppression is computed as the DQN reward for subsequent action generation. The DQN agent interacts continuously with the environment for training, and the detailed environment definition is given below:
%%%%%%%%%%%%%%%%%%%%%%%%%%%%%%%%%%%%%%%%%
\begin{algorithm}[t]
\SetAlgoLined
\LinesNumbered
\SetKwProg{Function}{function}{}{end}
\SetKwRepeat{Do}{do}{until}
% \begin{algorithmic}
\textbf{Require:} Reflectarray parameters, incident angle \eqr{(\theta_{inc},\varphi_{inc})}, and reflected angles \eqr{(\theta_{ref}^t,\varphi_{ref}^t)}, $\it{t={\rm 1},{\rm 2},\cdots,T}$

\textbf{Initialization:} Multi-focus profile \eqr{\textbf{P}_{mul}=0}, single-focus profiles \eqr{[\textbf{P}_{sing}^{1}, ... , \textbf{P}_{sing}^{T}]=0}, normalized~far-field intensity \eqr{\textbf{E}_{mul}=0}

\For{\eqr{t=1,...,T}}{

Calculate \eqr{\textbf{P}_{sing}^{t}} base on \eqr{(\theta_{inc},\varphi_{inc})} and \eqr{(\theta_{ref}^t,\varphi_{ref}^t)}

}

Calculate \eqr{\textbf{P}_{mul}} base on \eqr{[\textbf{P}_{sing}^{1}, ... , \textbf{P}_{sing}^{T}]}

Calculate \eqr{\textbf{E}_{mul}} (Eq. 5) base on \eqr{\textbf{P}_{mul}}

Find sidelobes \eqr{(\theta_{side}^m,\varphi_{side}^m)}, $\it{m={\rm 1},{\rm 2},\cdots,M}$ from \eqr{\textbf{E}_{mul}}

\For{\eqr{m=1,...,M}}{

Calculate \eqr{\textbf{P}_{side}^{m}} base on \eqr{(\theta_{inc},\varphi_{inc})} and \eqr{(\theta_{side}^m,\varphi_{side}^m)}

}

Deep Q-Network environment setup

\While{not Done}{

Reset the environment, set the initial state \eqr{s=s_0}

\For{\eqr{m=1,...,M}}{

Generate action \eqr{a} from action space \eqr{A}

Calculate \eqr{[\delta_m, \xi_m]} base on \eqr{a}

Remove \eqr{\delta_m \times \xi_m \times \textbf{P}_{side}^{m}} from \eqr{\textbf{P}_{mul}}

}

Calculate \eqr{\textbf{E}_{mul}^{*}} base on \eqr{\textbf{P}_{mul}}

Find \eqr{E_{intended}} (Eq. 6) and \eqr{E_{unwanted}} (Eq. 7) from \eqr{\textbf{E}_{mul}^{*}}

Reward \eqr{r=E_{intended}-E_{unwanted}}

}

\caption{Proposed sidelobe suppression method for multi-focus reconfigurable reflectarray.}
\label{al:DRL}
\end{algorithm}
%%%%%%%%%%%%%%%%%%%%%%%%%%%%%%%%%%%%%%%%%
% \vspace{-1cm}
\begin{table}[t]
\centering
\renewcommand{\arraystretch}{0.15}% Tighter
\caption{Hyperparameters setting of DRL}
\vspace{-1.50mm}
\footnotesize
\label{table:Hyperparameters}
\begin{tabular}{ll} % {l|l}
\toprule
Name & Value\\ \midrule
RL algorithm & DQN \\ 
% Exploration rate $\epsilon$ & Decay\\ 
Batch size & 128\\ 
Target network update interval & 100 \\ 
Reward discount factor $\gamma$ & 0.98 \\ 
Optimizer & Adam \\ 
Learning rate & 0.01 \\ 
NN type & Fully connected network \\ 
Number of neurons of each layer & [200,100,40] \\ 
Activate function (not for output layer) & Relu \\ 
Activate function for output layer & Linear \\ 
\bottomrule
\end{tabular}
\renewcommand{\arraystretch}{1.00}% Back to normal
\vspace{-4.50mm}
\end{table}
\begin{itemize}[leftmargin=*]
\item \textit{Action:} We utilize $\delta$ and $\xi$ to constitute a rectangular coordination system, forming the action space $\mathcal{A}$. Each point indicates a combination coefficient of specific $\delta$ and $\xi$ values. The number of steps in each iteration corresponds to the beams to suppress. During each step, the DQN agent generates a new combined coefficient based on exploration and exploitation principles, assigning the associated $\delta$ and $\xi$ to the targeted beam in that particular step.
% We utilize the parameter $\delta$ and $\xi$ sampling ranges as the x- and y-axis to form a rectangular coordinate system. Each point in this system indicates a combination coefficient defined by a pair of $\delta$ and $\xi$ values. The action space $\mathcal{A}$ consists of these combination coefficients covering all possible $\delta$ and $\xi$ values. Each iteration encompasses multiple steps, equivalent to the number of beams requiring suppression. During each step, the DQN agent generates a new combined coefficient based on exploration and exploitation principles. Then, the associated $\delta$ and $\xi$ are assigned to the beam targeted for suppression in that particular step.

\item \textit{State:} The environment state $\mathcal{S}$ is a vector with dimension of [Number of intended beams + Number of suppressed beams $\times$ 2]. It comprises the highest far-field intensity for each intended and suppressed beam, along with a combination coefficient indicating the parameters $\delta$ and $\xi$ for each suppressed beam.

\item \textit{Reward:} The reward $\mathcal{R}$ is defined as the difference~(in dB) between the minimum intensity of intended beams and the maximum intensity of unwanted sidelobes.

\item \textit{Hyperparameters:}
The DQN hyperparameters employed here are detailed in Table \ref{table:Hyperparameters}. These values are grounded in our practical experience gained from numerous simulations with diverse settings. Notably, the model training employs an exploration decay scheme linked to the iteration time.
% \begin{equation}
% \epsilon^*=\left\{
% \begin{array}{ccl}
% 0 & & {0\leq t\leq 100}\\
% \epsilon & & {100 \le t\leq 200}\\
% \epsilon+\frac{t-200}{4000} & & {200 \le t\leq 1000}\\
% 1 & & {t\ge 1000}
% \end{array} \right.
% \label{eq:adaptive decay}
% \end{equation}
\end{itemize}

%%%%%%%%%%%%%%%%%%%%%%%%%%%%%%%%%%%%%%%%%%%%%%%%%%%%%%%%%%%%%%%%%%%%%%%%%%%%%
\section{Result and Discussion}
% \vspace{-1.00mm}
\label{sec:Result and Discussion}
\subsection{Simulation Results}

\begin{figure}[t]
\centering
\includegraphics[width=50mm]{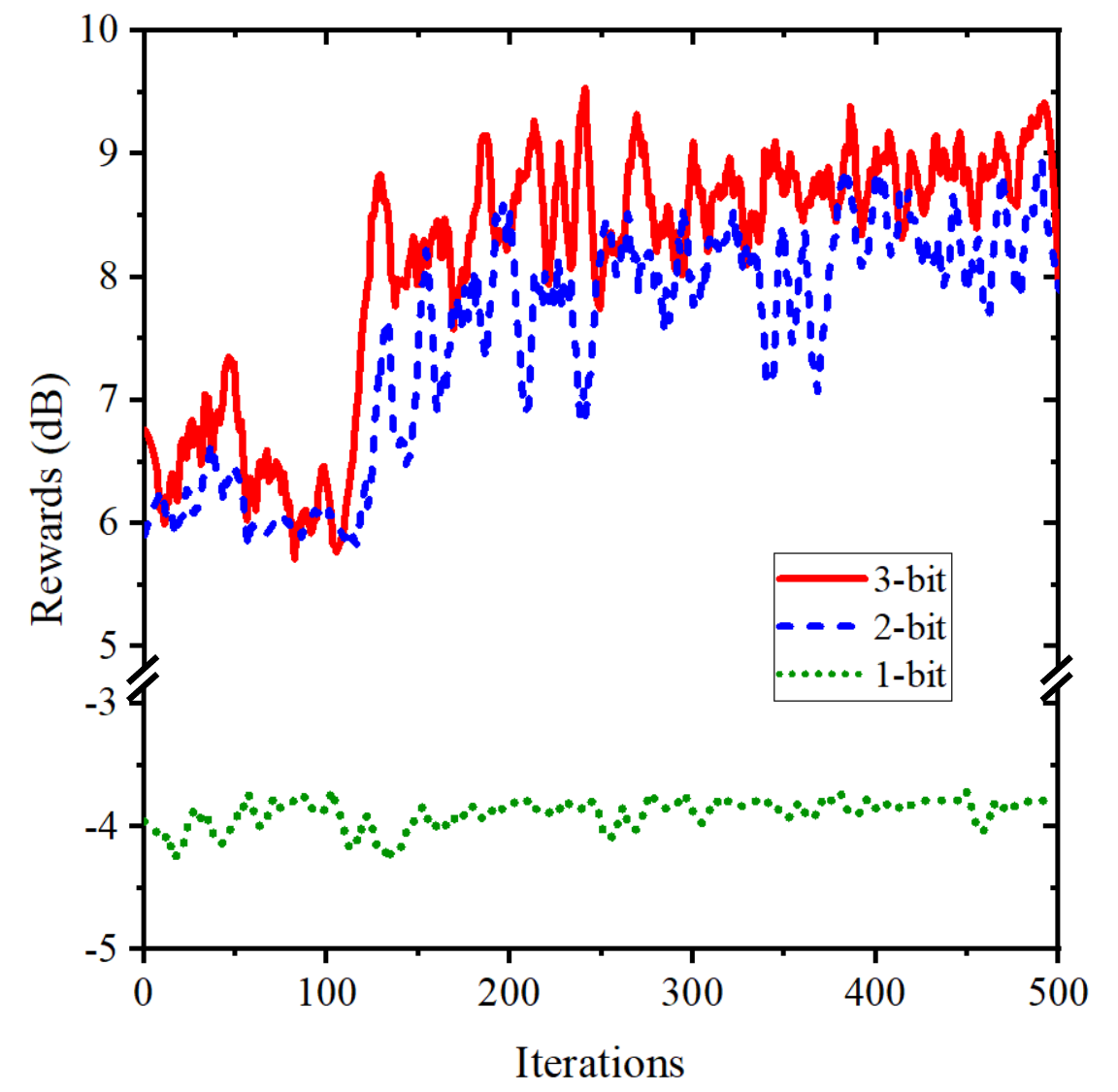}
\vspace{-3.25mm}
\caption{The smoothed DRL training curve with the RRA resolution of 1-bit, 2-bit, and 3-bit.}
\label{fig:curve}
\vspace{-4.50mm}
\end{figure}

\begin{figure*}[t]
\centering
\includegraphics[width=140mm]{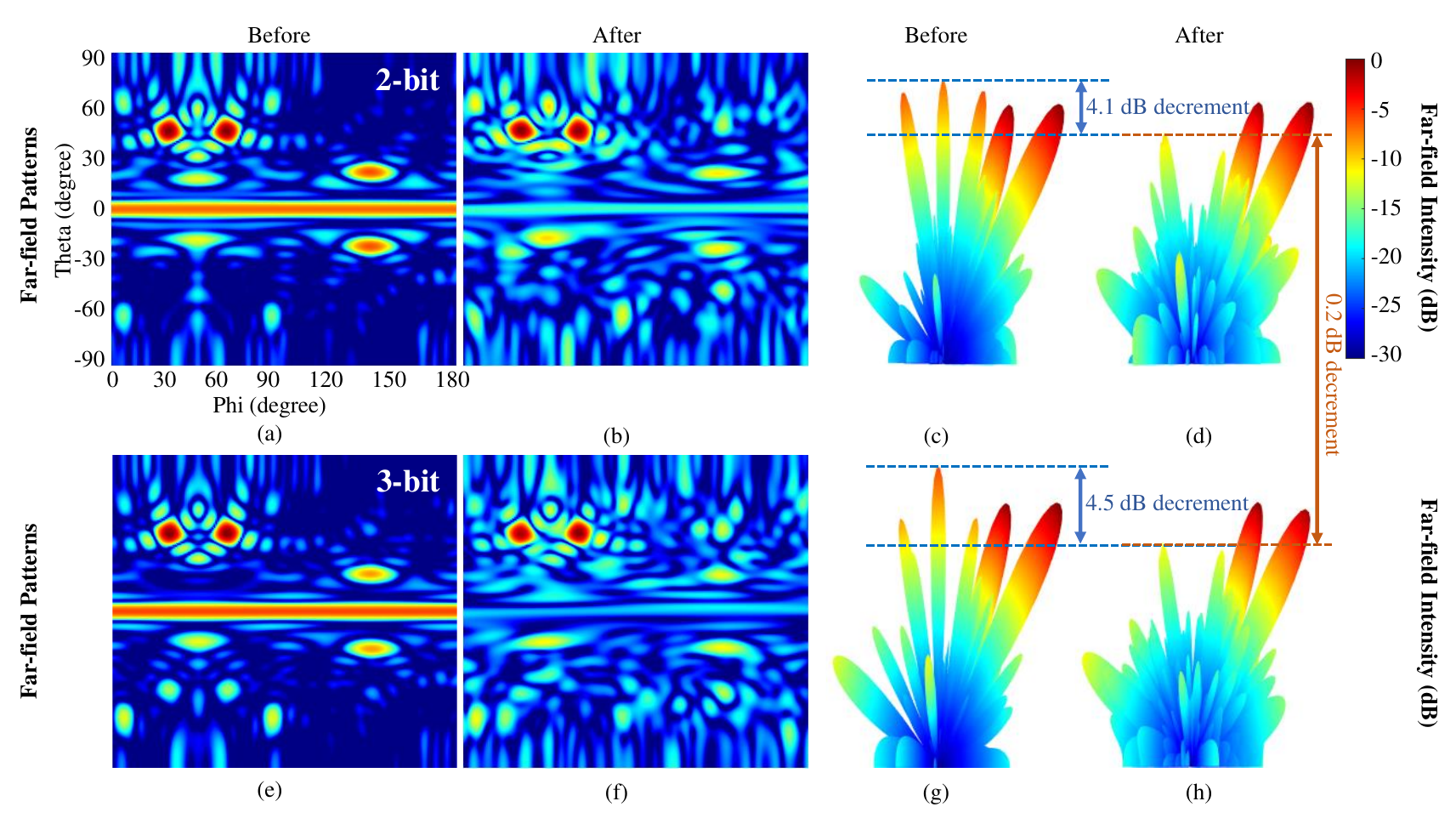}
\vspace{-4.50mm}
\caption{The far-field patterns (a)(c)(e)(g) before and (b)(d)(f)(h) after the proposed DRL algorithm. (a-d) is the result when RRA has a 2-bit resolution and (e-h) is when RRA has a 3-bit resolution.
The far-field intensities have been normalized to the maximum value of the intended beams after DRL training.}
\label{fig:performance}
\vspace{-4.50mm}
\end{figure*}

Taking the $50\times50$ RRA pointing at (45\degree,~30\degree) and (45\degree,~60\degree) as an example that works at 3.5GHz with a 17~mm distance (around 1/5~wavelength) between the center point of every two adjacent elements.
Fig.~\ref{fig:curve} illustrates the DRL training curve for RRA resolutions of 1-bit, 2-bit, and 3-bit (higher resolutions can be extended if necessary). Notably, the 1-bit resolution RRA exhibits unsatisfactory convergence, with unwanted lobes even stronger than intended beams in far-field validation. It is due to the inherent limitation of 1-bit resolution for only having two quantized reflect phases, leading to a mirrored reflection beam with the same intensity for single-focus far-field reflection. Therefore, errors from too-low resolution are further compounded in multi-focus applications. For resolutions higher than 1-bit, both the original and optimized performance improve with the increment of RRA resolution.
Because of introducing random parameter $\delta$, the training curves in Fig.~\ref{fig:curve} have been smoothed to make the training tendency more obvious. Nevertheless, a peak optimized gain of around 4 dB is still achieved compared with the performance before DRL.

Comparisons between far-field patterns before and after applying the proposed DRL algorithm to RRA with 2-bit and 3-bit resolutions are shown in Fig.~\ref{fig:performance}. The results show a 4.1 dB increment in sidelobe suppression with 2-bit resolution and a 4.5 dB improvement with 3-bit resolution. This raises the intensity difference between intended and undesired beams from approximately 6 dB to over 10 dB. Despite the slightly higher initial intensity of the 3-bit resolution, the maximum sidelobe intensity post-DRL remains 0.2 dB lower than that of the 2-bit resolution.

\subsection{Discussion and Future Works}
This paper aims to offer a time-efficient optimization solution for RIS. Hence, the DRL agent should be pre-trained offline and deployed online for lower inference time.
Through incorporating the randomizing parameter $\delta$ and suppression parameter $\xi$, the reflecting tendencies of undesired directions are eliminated and the periodicities are broken up from the multi-reflection profile. However, $\delta$ also brings about a series of fluctuations on the training curve because of the randomization for which further investigation is required in the future.

\section{Conclusion}
\label{sec:Conclusion}

This study addresses the issue of undesired sidelobes in the far-field pattern of multi-focus reconfigurable reflectarray. Leveraging the widely employed superposition principle for generating multi-focus reflections in RIS, we introduce a method that efficiently controls a pair of parameters to suppress these undesired sidelobes. A DRL-based algorithm is employed to pinpoint the optimal parameters. The performance of the proposed method has an improvement of 4 dB on average and achieves an obvious difference between the intended beams and undesired beams of around 10 dB, which means the sidelobe in a multi-focus reflection RIS system is suppressed significantly, and so to the power and information leakage.

%%%%%%%%%%%%%%%%%%%%%%%%%%%%%%%%%%%%%%%%%%%%%%%%%%%%%%%%%%%%%%%%%%%%%%%%%%%%%

% \vspace{-1.00mm}
\section*{Acknowledgments}
% \vspace{-1.00mm}
This work was developed within the Innovate UK/CELTIC-NEXT European collaborative research and development project on AIMM (AI-enabled Massive MIMO), and was partially supported by China Scholarship Council for funding the author under No.202108060224.
% \vspace{-2.50mm}
\bibliographystyle{IEEEtran} %
\balance

\bibliography{IEEEabrv,references} 
% \vspace{-2.50mm}
% \clearpage

\end{document}